\documentclass[10pt,twocolumn]{article}

\usepackage[letterpaper,margin=0.8in]{geometry}
\usepackage{hyperref}
\usepackage{natbib}

\usepackage{astrojournals}
\usepackage{graphicx}
\usepackage{mathtools}
\usepackage{amsmath}
\usepackage{abstract}
\usepackage{lipsum}
\usepackage[charter]{mathdesign}
\usepackage{microtype}
\usepackage{xspace}
\usepackage{xcolor}
\usepackage{paralist}
\usepackage{enumitem}
\usepackage{float}
\usepackage{caption}
\usepackage{sectsty}
\allsectionsfont{\raggedright\bfseries\large}
\subsectionfont{\raggedright\normalfont\itshape\normalsize}

\def\icarus{\ref@jnl{ICARUS}}             

\newcommand*{\note}[1]{{{{#1}}}}

\newcommand*{\fhelper}[1]{\ensuremath{\vec{f}_{\!\text{#1}}}\xspace}
\newcommand*{\ihelper}[1]{\ensuremath{i_{\!\text{#1}}}\xspace}

\newcommand*{\fr}{\fhelper{r}}

\newcommand*{\ie}{\ihelper{e}}
\newcommand{\Md}{M_{\rm disk}}
\newcommand{\Mbh}{M_{\rm \bullet}}

\newcommand{\Mo}{M_{\odot}}
\def\arcsec{\hbox{$^{\prime\prime}$}}

\newcommand{\eg}{e.\,g.\xspace}
\renewcommand{\deg}{\ensuremath{^{\circ}}\xspace}
\newcommand{\peryr}{{\rm yr}^{-1}\xspace}
\newcommand{\pergal}{{\rm gal}^{-1}\xspace}

\renewcommand*{\vec}[1]{\boldsymbol{#1}}

\let\oldhat\hat
\renewcommand*{\hat}[1]{\vec{\oldhat{#1}}}

\hypersetup{%
 pdftitle={TDEs},
 pdfauthor={\textcopyright\ authors},
 bookmarksopen=true,
 colorlinks=true,
 linkcolor=black,
 citecolor=black,
 urlcolor=black}

\defcitealias{Madigan2017}{M17}

\usepackage{etoolbox}

\makeatletter
\let\@fnsymbol\@arabic
\makeatother

\makeatletter

\patchcmd{\NAT@citex}
  {\@citea\NAT@hyper@{\NAT@nmfmt{\NAT@nm}\NAT@date}}
  {\@citea\NAT@nmfmt{\NAT@nm}\NAT@hyper@{\NAT@date}}
  {}
  {}

\patchcmd{\NAT@citex}
  {\@citea\NAT@hyper@{%
     \NAT@nmfmt{\NAT@nm}%
     \hyper@natlinkbreak{\NAT@aysep\NAT@spacechar}{\@citeb\@extra@b@citeb}%
     \NAT@date}}
  {\@citea\NAT@nmfmt{\NAT@nm}%
   \NAT@aysep\NAT@spacechar%
   \NAT@hyper@{\NAT@date}}
  {}
  {}

\patchcmd{\NAT@citex}
  {\@citea\NAT@hyper@{%
     \NAT@nmfmt{\NAT@nm}%
     \hyper@natlinkbreak{\NAT@spacechar\NAT@@open\if*#1*\else#1\NAT@spacechar\fi}%
       {\@citeb\@extra@b@citeb}%
     \NAT@date}}
  {\@citea\NAT@nmfmt{\NAT@nm}%
   \NAT@spacechar\NAT@@open\if*#1*\else#1\NAT@spacechar\fi%
   \NAT@hyper@{\NAT@date}}
  {}
  {}

\makeatother

\begin{document}
\fontsize{10}{14}\selectfont

\title{Dynamical Properties of Eccentric Nuclear Disks: Stability, Longevity, and Implications for Tidal Disruption Rates in Post-Merger Galaxies}

\author{Ann-Marie Madigan,%
\thanks{Astrophysical and Planetary Sciences \& JILA, \newline 
CU Boulder, CO; { annmarie.madigan@colorado.edu} } \
Andrew Halle,%
\thanks{UC Berkeley, CA} \
Mackenzie Moody,%
\thanks{Princeton University, NJ} \
Michael McCourt,%
\thanks{UC Santa Barbara, CA; {NASA Hubble Postdoctoral Fellow}} \
Chris Nixon,%
\thanks{Theoretical Astrophysics Group, Department of Physics \& Astronomy, University of Leicester} \\
\note{\& Heather Wernke}$^1$ 
}

\date{\today}

\twocolumn[
\maketitle
\begin{onecolabstract}

In some galaxies, the stars orbiting the supermassive black hole take the form of an eccentric nuclear disk, in which every star is on a coherent, apsidally-aligned orbit. The most famous example of an eccentric nuclear disk is the double nucleus of Andromeda, and there is strong evidence for many more in the local universe. Despite their apparent ubiquity however, a dynamical explanation for their longevity has remained a mystery: differential precession should wipe out large-scale apsidal-alignment on a short timescale.

Here we identify a new dynamical mechanism which stabilizes eccentric nuclear disks, and explain for first time the negative eccentricity gradient seen in the Andromeda nucleus. 
The stabilizing mechanism drives oscillations of the eccentricity vectors of individual orbits, both in direction (about the mean body of the disk) and in magnitude. Combined with the negative eccentricity gradient, the eccentricity oscillations push some stars near the inner edge of the disk extremely close to the black hole, potentially leading to tidal disruption events. 

Order of magnitude calculations predict extremely high rates in \note{recently-formed eccentric nuclear disks} ($\sim0.1 - 1$ $\peryr \pergal$). 
Unless the stellar disks are replenished, these rates should decrease with time as the disk depletes in mass. \note{If eccentric nuclear disks form during gas-rich major mergers,} this may explain the preferential occurrence of tidal disruption events in recently-merged and post-merger (E+A/K+A) galaxies.

\bigskip
\end{onecolabstract}
\vspace*{2\baselineskip}
]
\saythanks

\raggedbottom
\section{Introduction}
\label{sec:intro}
The tidal gravity of supermassive black holes (SMBHs) gives rise to some of the
most energetic phenomena in the universe: tidal disruption events,
hyper-velocity stars, and gravitational wave inspirals of compact  
stellar remnants.  Fundamentally, the rates of these events are determined by the dynamics of stars in the vicinity of SMBHs.  
It is commonly assumed that nuclear star clusters are spherically symmetric, and that gravitational two-body
scattering is the dominant process determining the rates of these
events.  
However, the distribution of stars around SMBHs is often observed to be highly asymmetric, and coherent torques between stellar orbits may strongly dominate over two-body interactions. 
For a
historical perspective, we look to our most massive
galactic neighbor.

The asymmetric nucleus of the Andromeda galaxy (M31) has been a puzzle
since its discovery by balloon-borne experiments \citep{Light1974}. 
It took nearly two decades for a telescope of sufficient resolution (\textit{Hubble Space Telescope}; HST) to resolve the nucleus into two distinct components: a faint peak (P2) lying approximately at the bulge center, and a brighter component (P1) offset by $0.\arcsec5$ \citep[$\approx 2$ pc;][]{Lauer1993}.  
An obvious hypothesis is that P1 and P2 are separate star
clusters. If this were the case, however, the two clusters should
merge on such a short timescale that finding something like the M31
nucleus becomes exceedingly improbable. Furthermore, spectroscopy reveals that their K-type stellar populations are nearly identical \citep{Kormendy1999}; they must therefore comprise a single system of stars. 

\citet{Tremaine1995} showed that the double nucleus is best modeled as an eccentric stellar disk orbiting the SMBH. This configuration can lead to two brightness peaks: a primary peak at apoapsis (P1), where stars spend most of their time, and a fainter one at periapsis (P2), where orbits pinch together and create an enhancement in surface brightness. The periapsis peak is faint, however, and is only visible for a narrow range of nearly edge-on orientations. 
Though the M31 disk at first seems exotic, seeing an improbable arrangement in a nearby galaxy suggests that these disks should in fact be common.
Indeed, while no systematic search for eccentric nuclear stellar disks has been conducted, there is growing evidence that they are unexceptional. 
In the \citet{Lauer2005} sample of 65 non-dust-obscured early-type galaxies, $\sim$20\% show features consistent with eccentric nuclear stellar disks seen from different angles (offset nuclei, nuclei with central minima and double nuclei). 
This is despite the observational challenges of detecting such signatures; stable eccentric nuclear disks live within the radius of influence of the SMBH, a scale difficult to resolve except for the closest and/or most massive galaxies (see \S~\ref{sec:disc}). 

\citet{Hop10a,Hop10b} show that eccentric nuclear stellar disks may originate in gas-rich galaxy mergers: they form when gas that is funneled to the center of the potential fragments and forms stars on aligned eccentric orbits. The presence and prevalence of these disks may therefore  encode a wealth of information about galaxy formation and merger histories. 
Despite the importance and apparent ubiquity of eccentric nuclear disks however, the stability of such disks has remained a mystery. 
Individual orbits within the disk should precess at different rates, smearing out the coherent structure on a differential precession timescale of $\sim$Myr, far shorter than the $\gtrsim$ Gyr ages of the stars in the M31 disk. This has led many \citep[e.g.,][]{Merritt2013} to conclude that the eccentric disk we see in M31 is a transient feature. 

In this paper we identify a new dynamical mechanism which
  stabilizes eccentric nuclear stellar disks against smearing by differential precession, thus explaining the
longevity of these structures.  
We show that in an isolated system, \note{a small fraction of outermost orbits break away to form their own oppositely-precessing structure}, for which there may be evidence in the M31 nucleus \citep{Men13,Bro13}. 
We also show that the mechanism naturally reproduces the
  negative eccentricity gradient of the M31 disk \citep{Pei03}. 
An important corollary of our disk stabilizing mechanism is that the stars undergo oscillations in orbital eccentricity. 
During these oscillations, some fraction of the stars approach plunging orbits and are susceptible to tidal disruption by the SMBH near periapsis.
Thus, {we propose that eccentric nuclear disks, via their stabilizing mechanism,
 fundamentally change the rates at which stars interact with
 SMBHs.} Such interactions may significantly
 enhance the production rates of tidal disruption events,
 hyper-velocity stars and binaries, and gravitational wave inspirals
 of compact objects. 
 
 We present the paper as follows: 
 in \S~\ref{sec:model} we describe a new dynamical model which explains the stability of eccentric nuclear disks. In \S~\ref{sec:predictions} we compare predictions of our model with $N$-body results, showing that an eccentric disk develops a negative eccentricity gradient, and that individual orbits undergo large-amplitude oscillations in eccentricity and that some may even flip in inclination.  In \S~\ref{sec:tdes} we describe how our dynamical model can explain the enhanced tidal disruption events seen in recently-merged and post-merger galaxies. 
 In \S~\ref{sec:disc} we summarize our findings and discuss our results. 

\begin{figure*}
  \centering
  \includegraphics[width=0.9\textwidth]{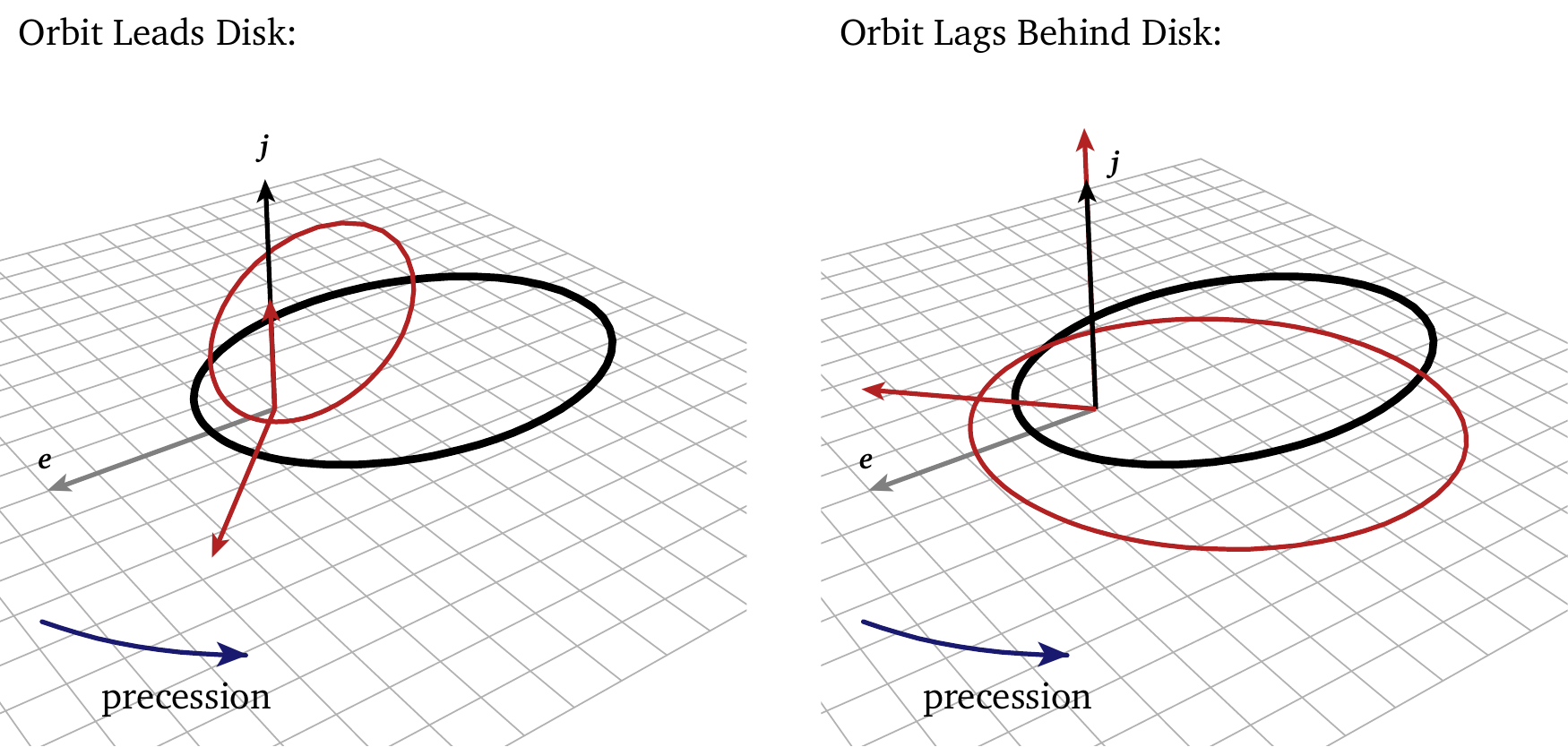}
\caption{\textbf {Physics of orbital oscillations within a stable eccentric disk.} \\*[0.5\baselineskip]
The entire disk (black) precesses in the prograde direction (counter-clockwise in this figure). If an orbit (red) moves ahead of the disk (left panel), it feels a gravitational pull towards the bulk of the disk. This torques the orbit, decreasing its angular momentum and thus increasing its orbital eccentricity ($j^2 \propto (1 - e^2)$). This lowers the orbit's precession rate, allowing the bulk of the disk to catch up with it. The reverse happens for an orbit which lags behind the bulk of the disk (right panel). The overall effect is to stabilize the disk: any orbit which is perturbed off the disk is driven back toward it by torques and differential precession. The mechanism inducing this stability leads to oscillations in eccentricity. }
 \label{fig:model}
\end{figure*}

\section{A Stability Model for Eccentric Nuclear Disks}
\label{sec:model}

The orbit of a star near a SMBH of mass $M_{\bullet}$ can be described by two vectors: the (specific) angular momentum vector ${\vec{j}} = \vec{r} \times \vec{v}$, which defines the orbital plane, and the eccentricity vector $\vec{e} = (\vec{v} \times \vec{j})/ (G M_{\bullet}) - \hat{r}$, which points toward periapsis and whose magnitude equals the eccentricity of the orbit. 
In a perfectly Keplerian potential, the vectors $\vec{j}$ and $\vec{e}$ are constant. If we add a small non-Keplerian force $\vec{f}$, however, the orbit vectors evolve in time according to 
\begin{subequations}
\begin{align}
{\vec{j}}^\prime &= \vec{r} \times \vec{f} = \vec{\tau}\label{eq:1} \\
{\vec{e}}^\prime &= \frac{\vec{f} \times \vec{j}}{G M_{\bullet}} + \frac{\vec{v} \times \vec{\tau}}{G M_{\bullet}}, & & \label{eq:2}
\end{align}
\end{subequations}
where $\vec{\tau} \equiv \vec{r} \times \vec{f}$ is the specific torque produced by the non-Keplerian force. 

An eccentric disk is characterized by apsidally-aligned orbits, i.e., orbits with aligned $\vec{e}$-vectors. Non-Keplerian forces induce precession of the $\vec{e}$-vectors (equation~\ref{eq:2}) at rates that depend on the semi-major axes of the orbits (orbits near the inner edge of the disk have smaller angular momentum and experience larger forces than orbits at the outer, less dense edge of the disk). Hence we would expect the $\vec{e}$-vectors to spread out on a differential precession timescale, and that the apsidally-aligned orbits observed in the M31 nucleus and in so many other galaxies should be a short-lived, transient phenomenon.  

The forces that drive differential precession in $\vec{e}$-vectors, however, also result in coherent gravitational torques (equation~\ref{eq:1}) acting between stellar orbits \citep{Rauch1996}.
In the inner arcsecond ($\approx 4$ pc) of the M31 nucleus, where the eccentric nuclear stellar disk can be studied in detail, the total mass of the disk is $\sim 15\%$ of the mass of the SMBH \citep[$\Mbh \approx 1-2 \times 10^8 \Mo$;][]{Bender2005}. Hence the secular (orbit-averaged) gravitational torques exerted by the disk orbits on each other are strong and can effectively counteract $\vec{e}$-vector differential precession.  

The stability of a disk depends entirely on the direction of its $\vec{e}$-vector precession. 
In \citet{Madigan2009}, we focused on the case in which an eccentric stellar disk is embedded in a more massive $\sim$symmetric nuclear star cluster \citep[such as in the Milky Way Galactic center;][]{Feldmeier2014}. This additional gravitational potential leads to retrograde precession\footnote{
An intuitive explanation for this retrograde precession can be seen from equation~\ref{eq:2}.  In Gauss's approximation, we spread the mass of the star over its orbit, with the density inversely proportional to the instantaneous velocity. Hence most of the mass of the orbit is located at apoapsis. For sufficiently eccentric orbits and for typical nuclear star cluster density profiles, the forces experienced at this location dominate over those at periapsis. A spherical gravitational potential results in an inward radial force at apoapsis; the first term in equation~\ref{eq:2} gives retrograde precession.} of the orbits (${\vec{e}}^\prime \cdot \vec{v_p} < 0$ where $\vec{v_p}$ is the velocity at periapsis).
Retrograde precession, combined with mutual gravitational torques, results in an `eccentric disk instability' which propels the orbits apart. 

Here we focus on the case in which the mass of the \textit{asymmetric} eccentric disk is much greater than the background stellar potential \citep[as is true for the M31 nucleus;][]{Kormendy1999}, such that the direction of precession is reversed. 
Prograde precession (${\vec{e}}^\prime \cdot \vec{v_p} > 0$) leads instead to stability: 
orbits which precess ahead of the disk feel a gravitational pull toward the disk behind it. We show this in the left panel of figure~\ref{fig:model}. This gravitational force creates a torque $\tau_z < 0$ (see equation~\ref{eq:1}) which decreases the angular momentum of the orbit. 
Specific angular momentum and energy are defined as
\begin{subequations}
\begin{align}
j^2 &=  G\Mbh a (1 - e^2) \label{eq:j} \\
E &= \frac{G\Mbh}{2 a}. & & \label{eq:E}
\end{align}
\end{subequations}
The torque does not affect the energy, or equivalently the semi-major axis, of the orbit. Hence the torque raises the orbital eccentricity of the orbit. 
Increasing the eccentricity slows its angular precession rate ($\propto \vec{f} \times \vec{j}/e$), stalling the orbit until it is reabsorbed by the mean body of the disk. A similar analysis shows that orbits which lag behind the disk decrease in eccentricity, precess more rapidly and are driven back towards the bulk of the disk (right panel of figure~\ref{fig:model}). 
This stability mechanism implies both that a coherent precessing eccentric disk maintains its shape in response to perturbations, and that perturbed orbits undergo oscillations in eccentricity and in orientation about the mean body of the disk. 
A similar analysis shows that sufficiently massive eccentric disks are stable to perturbations in inclination resulting from out-of-plane forces.   
Strongly perturbed and/or extremely eccentric orbits can flip their orientation however; see \S~\ref{sec:flips}.

\section{$N$-body Simulations of Eccentric Disks}
\label{sec:predictions}

We can use these results to study stable models of eccentric disks using $N$-body simulations. 
We use the {\tt REBOUND} code with the IAS15 integrator \citep{Rein2012,Rein2015} with no gravitational softening. 
We chose general initial conditions for disk orbits, initializing $N = 100 - 1000$ stars on eccentric apsidally-aligned orbits with a narrow range of semi-major axes ($a = 1 - 2$), a flat eccentricity distribution, and surface density profile $\Sigma \propto a^{-2}$.  
The disk to SMBH mass ratio is $\Md/\Mbh = 0.01$. 
For simplicity, we treat the stars as point masses and do not include a tidal disruption radius in the code; orbits are allowed pass through $e = 1$. 

\begin{figure*}
  \centering
  \includegraphics[width=1.0\textwidth]{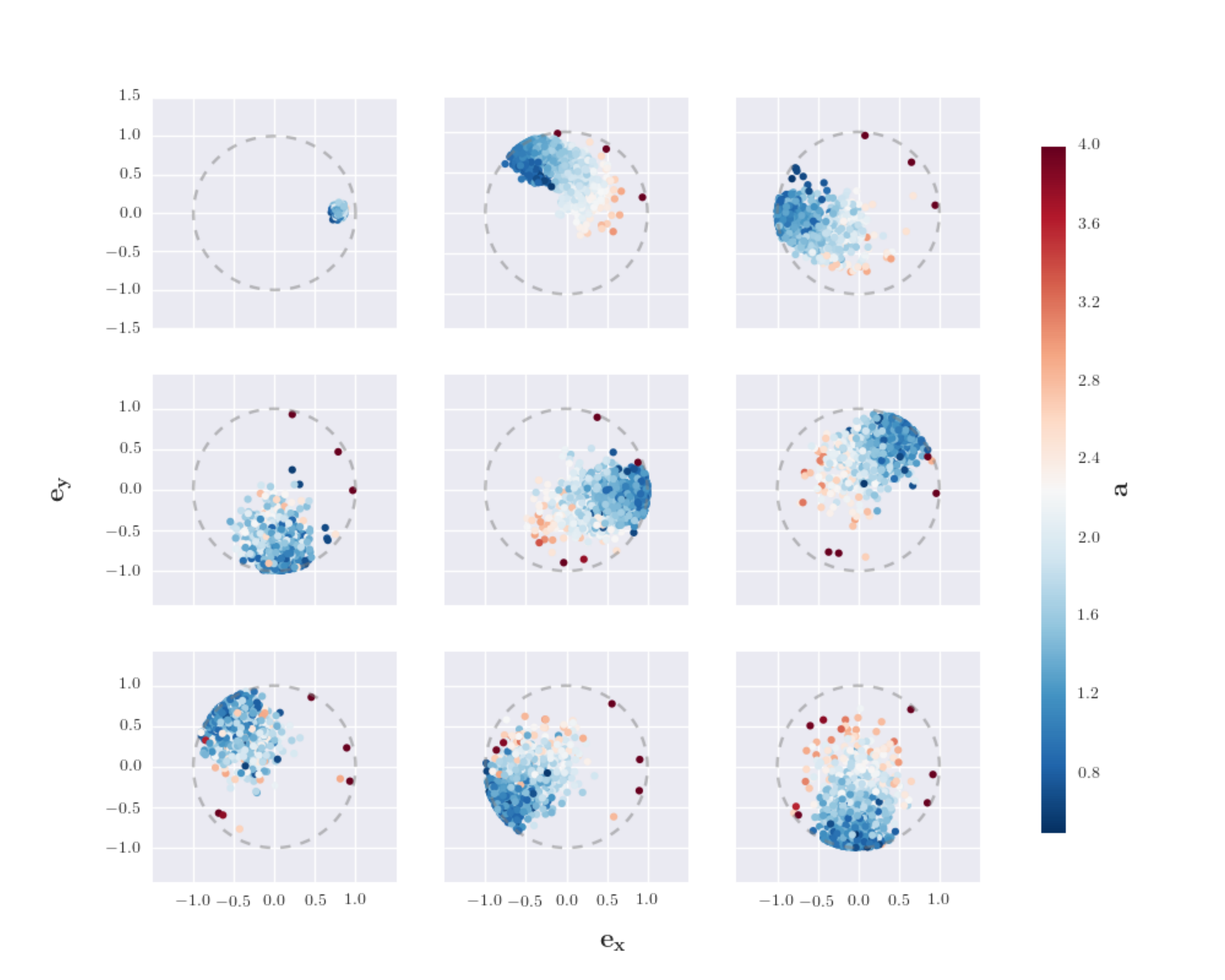}
\caption{\textbf {Time evolution of an eccentric disk} \\*[0.5\baselineskip]
Eccentricity vectors of stars in an $N$-body simulation (projected into the disk plane), plotted every secular timescale ($\sim 100$ orbits) (left to right, top to bottom). 
The dashed grey circle indicates values of $e = 1$. Stars are initialized with eccentricities $e = 0.8$. Colors correspond to semi-major axes (\note{stars are initialized from a = 1-2 but spread out quickly due to two-body relaxation}). 
The \note{bulk of the disk stably} precesses with prograde motion (counter-clockwise). 
The $\vec{e}$-vectors spread out due to orbits oscillating back and forth across the mean body of the disk and high eccentricity orbits gaining significant inclinations.  
\note{The very outermost orbits precess with retrograde motion (clockwise). 
They exchange angular momentum via coherent gravitational torques with the prograde-precessing disk when they encounter it at low relative azimuthal angles.}
 }
 \label{fig:evectors}
\end{figure*}

\subsection{Oppositely-precessing outer orbits}
\label{ss:outer-disk}

Our model posits that eccentric orbits in an asymmetric disk are stable so long as they apsidally precess in a prograde direction.  
We verify this with $N$-body simulations. Figure~\ref{fig:evectors} shows results for an eccentric disk with initial orbital eccentricities $e=0.8$. 
We plot the components of the eccentricity vectors in the disk plane $(e_x, e_y)$. 
\note{As expected, the bulk of the eccentric nuclear disk stably precesses with prograde motion (counterclockwise; blue points). The outermost orbits ($5-10\%$ by mass) are pushed via two-body relaxation beyond the initial outer edge of the disk, and precess more slowly with retrograde motion (clockwise; red points). The direction of precession of orbits at intermediate semi-major axes is time-dependent \--- these orbits undergo both prograde and retrograde precession as secular torques change their eccentricities. Dissipation decreases the mean precession rate of the disks.  
This is also due to two-body relaxation which increases the semi-major axis range of the stars as they scatter off one another.  }
 
 \begin{figure}
  \centering
  \includegraphics[width=1.0\columnwidth]{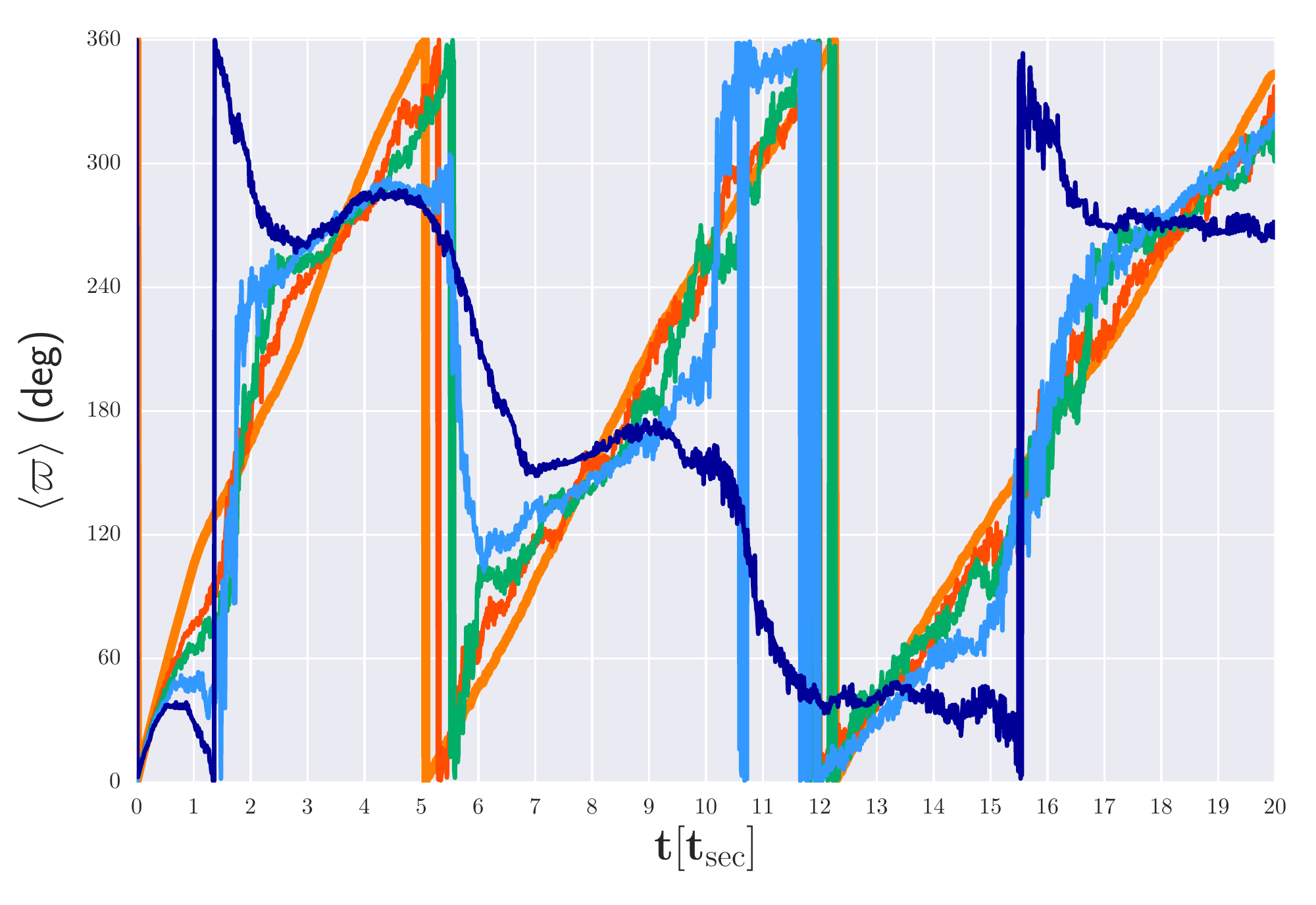}
  \caption{\textbf {Mean precession of disk orbits} \\*[0.5\baselineskip]
Mean longitude of periapsis, $\varpi$, as a function of time in units of the secular timescale, $t_{\rm sec} \equiv  {\Mbh}/{\Md}~ P$. The various colored lines show different sections of the disk binned in semi-major axis to highlight the reversal in the direction of precession of the outer orbits. 
The orange line shows $\langle\varpi\rangle$ for the innermost $76\%$ of the disk orbits (in semi-major axis). The red (76\% - 82\%), green (82\% - 88\%), and light blue (88\% - 94\%) lines show the precession for sections of the disk increasing in semi-major axis.
The dark blue line shows $\langle\varpi\rangle$ for the outermost $6\%$ of the disk orbits which have a net precession in the opposite, retrograde, direction. }
 \label{fig:opposite-structures}
\end{figure}

\note{
In figure \ref{fig:opposite-structures}, we show the transition between the prograde-precessing eccentric nuclear disk and the outermost retrograde-precessing orbits. The orange line shows the mean longitude of periapsis $\varpi$ as a function of time for the innermost $76\%$ of the disk orbits (in semi-major axis). The dark blue line shows the same for the outermost $6\%$ of the disk orbits.
Three things stand out from this plot. The first is that the transition between the inner prograding-precessing disk and the outer retrograde precessing orbits is not abrupt. Second, the outer orbits precess at about half the rate of the prograde-precessing disk. 
Third, the precession rate of the outer orbits slows when the orbits overlap with the main disk. 
This last point is important observationally: the two-oppositely precessing structures will not necessarily appear as separate components. 
The slow-down in precession is a direct result of angular momentum exchange. 
As the outer orbits approach the inner prograde-precessing disk, they loses angular momentum, increase in mean orbital eccentricity, and slow down in precession.  As they precesses past the disk, they are torqued to higher angular momenta, decrease in mean orbital eccentricity, and speed up in precession. Hence they appear to race around in precession towards the inner disk again. As the prograde-precessing disk is far more massive, it experiences smaller eccentricity variations. 
}

Our idealized, low-$N$ simulations can not provide a direct comparison to the M31 nucleus. 
Due to the artificially high two-body relaxation rate, the stars diffuse rapidly in semi-major axes thus decreasing both the orbital precession rates and the strength of the torques exerted between orbits. 
Both of these affect the eccentricity structure of the disk which feeds back into the precession rates. 
A careful convergence study will therefore be required to make a concrete comparison with the M31 nucleus. We must also add the M31 bulge potential and general relativistic precession, explore a smaller mass ratio between the SMBH and the disk, and expand the semi-major axis range of the disk. 
Nevertheless, across a broad range of parameter space, our simulations generically show that the outermost orbits ($\le 10\%$ in mass) break away into an oppositely-precessing structure, and there may be observational evidence of this in the M31 nucleus. 

\citet{Bro13} explore non-parametric models of the M31 nuclear disk. 
They find that some sections of the disk are not aligned with others. This is consistent with our model. 
In detail, however, they find that orbits are anti-aligned with respect to the main disk both inside $r < 0.\arcsec15$ and outside $r > 1\arcsec.2$. 
This is not what we predict, but this is likely an artifact of biaxial symmetry in their model.  
The inner structure that they describe behaves similarly to the inner disk in our simulations: eccentricities, inclinations, and their standard deviations decrease with semi-major axis. 
 There are differences in the main disk structure that they find however. The inclinations of orbits increase with radius which is the opposite of what we find. We note however that our simulations do not include a grainy ($N$-body) background potential; this has been shown to be important in the growth of orbital inclinations of disk stars \citep{Loc09b}. 

\citet{Men13} recently reported the discovery of an eccentric H$\alpha$-emitting disk in the inner $0.\arcsec7$ of the M31 nucleus.
The source of this emission -- whether stellar or gaseous in origin -- is not yet known. 
Using a simple orbit model, they derive Kepler elements that show it to be related to, but distinct from, the P1-P2 stellar disk. Its eccentricity, inclination and position angle of ascending nodes ($e = 0.35$, $i = 45\deg$, P. A. $= 59\deg$) are very similar to that derived for the stellar P1-P2 disk ($e = 0.3$, $i = 55\deg$, P. A. $= 56\deg$). Noticeably however, the argument of periapsis $\omega = 115\deg$ is very different from that of the stellar P1-P2 disk, $\omega = 250\deg$. This translates to a longitude of periapsis\footnote{$\varpi$ is the angle between the eccentricity vector and a reference direction.} difference of $\Delta \varpi \sim 135\deg $ between the two structures. 

{Though it is not yet clear how to map the \citeauthor{Bro13} models and \citeauthor{Men13} H$\alpha$-observations to our simulations or even to each other, \note{a stable eccentric nuclear disk is likely to be more complex than a single monolithic structure.}} 

\subsection{Negative eccentricity gradient of the main disk}
\label{ss:ecc-gradient}

\begin{figure}
  \centering
  \includegraphics[width=1.0\columnwidth, trim={0 4cm 3cm 0}, clip]{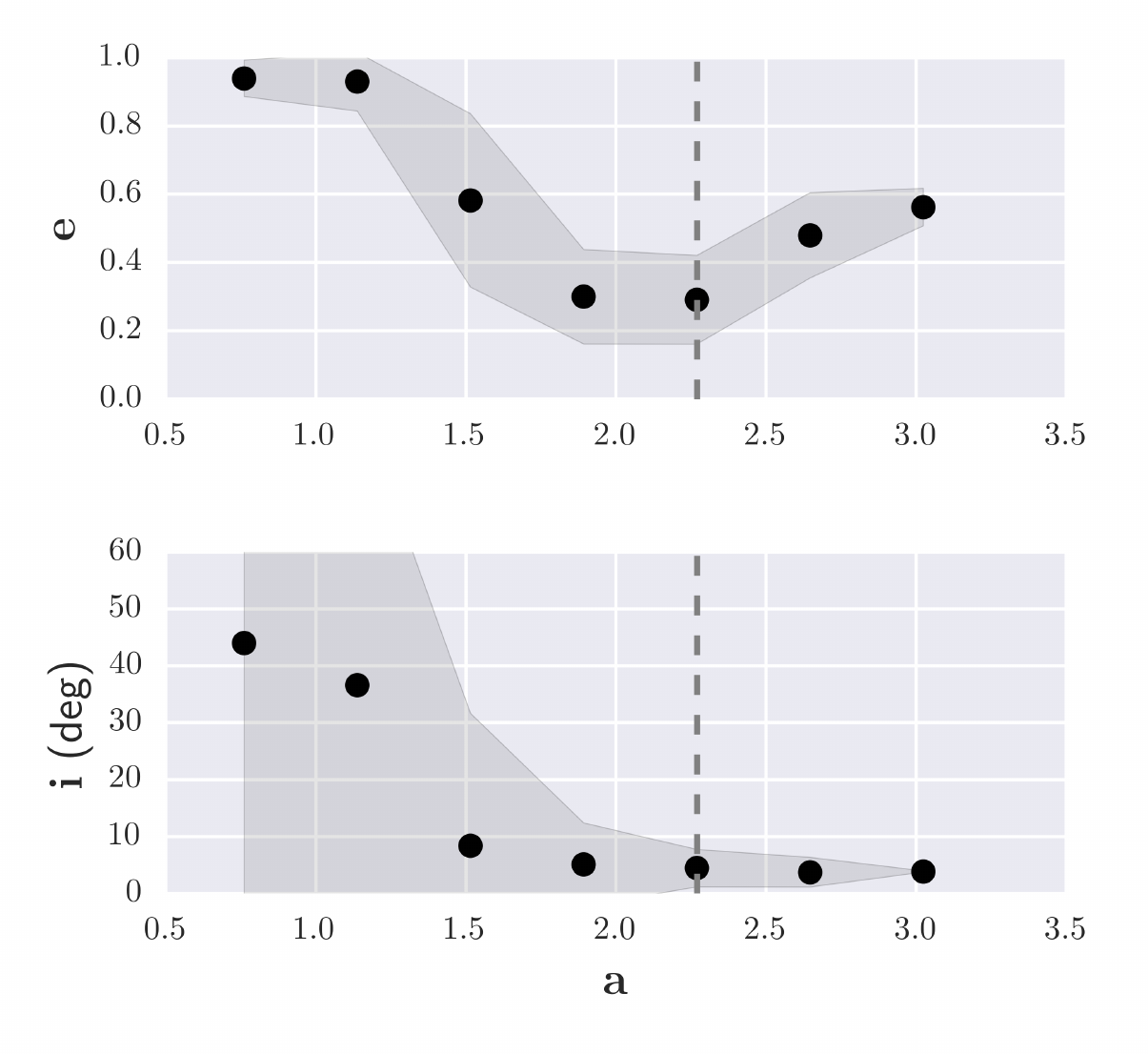}
\caption{\textbf {Structure of a stable eccentric nuclear disk} \\*[0.5\baselineskip]
\textit{Top}: eccentricity gradient as a function of semi-major axis after $\sim200$ orbits. 
Dots indicate the mean values of disk stars. 
The gray region shows one-sigma standard deviation.  
Stars at the innermost edge of the disk reach $e\sim1$, and are susceptible to disruption. At $t = 0$ the disk eccentricity is $e = 0.8$. 
\textit{Bottom}: inclination gradient as a function of semi-major axis.  Dots indicate the median values of disk stars. 
  \note{The vertical dashed lines indicate the division between the stable prograde-precessing disk and the outer retrograde-precessing orbits.}
}
 \label{fig:e_v_a}
\end{figure}

Orbits of the same eccentricity with low semi-major axes will precess faster than those with high semi-major axes (from equation~\ref{eq:2}; ${\vec{e}}^\prime  \propto {\vec{f} \times \vec{j}} \propto (1 - e^2)^{1/2} ~a^{-3/2}$ taking $f \propto a^{-2}$).  For the inner disk to stably precess as one body, the orbits at low semi-major axes must increase their eccentricities with respect to the mean, while the orbits at high semi-major axes must become more circular. They do this naturally via the coherent torques as described in \S~\ref{sec:model}. Hence our model predicts that a stable equilibrium eccentric stellar disk will have a negative eccentricity gradient: orbital eccentricities decrease as a function of semi-major axis.  

Our $N$-body simulations verify this prediction. As an example we show results from a simulation with initial disk eccentricity $e = 0.8$.
The top panel of figure~\ref{fig:e_v_a} shows the negative eccentricity gradient $de/da < 0$ of the inner disk, with a high mean eccentricity $\langle e \rangle \approx 0.95$ near its inner edge. The grey regions show one-sigma quantiles, the typical amplitude of eccentricity oscillations (see \S~\ref{ss:oscillations}). 
The outer edge of the inner disk has a much lower mean eccentricity of $\langle e \rangle \approx  0.3$. 
Beyond $a \gtrsim 2.3$ \note{the orbits precess with retrograde motion}.
The bottom panel of figure~\ref{fig:e_v_a} shows the inclination distribution as a function of semi-major axis, which also demonstrates a clear negative gradient, $di/da < 0$.   
The highest values at low semi-major axes are driven by the `flipping' of extreme eccentricity orbits to $i \sim 180\deg$ (see \S~\ref{sec:flips}). 
In this high eccentricity disk, secular (orbit-averaged) torques dominate the dynamical evolution of the system. Orbits with high eccentricities have less angular momentum and therefore require less force to be torqued around, and so a negative inclination gradient is a natural consequence of the negative eccentricity gradient. 

One might expect from the inclination distribution that the inner disk is more geometrically thick than the outer disk. This is not the case however, as pointed out by \citet{Haas2016}: the high eccentricity orbits have $\omega$ values such that they are relatively embedded in the disk.  In other words, inclinations grow via orbits rolling over their major axes. Since the orbits are so eccentric, their high inclinations do not lift them far above the disk in 3D space.
In terms of the azimuthal distribution of orbits, the \note{stable eccentric nuclear disk} is a coherent, prograde-precessing structure with a relatively small deviation of $\varpi$ values, whereas the outermost orbits form a less coherent structure.  
The full three-dimensional structure of eccentric disks, which varies with semi-major axis and also with azimuth, will greatly affect the observational signatures of disks seen from different angles.

\subsection{Eccentricity oscillations}
\label{ss:oscillations}
\begin{figure}
  \centering
  \includegraphics[width=1.0\columnwidth]{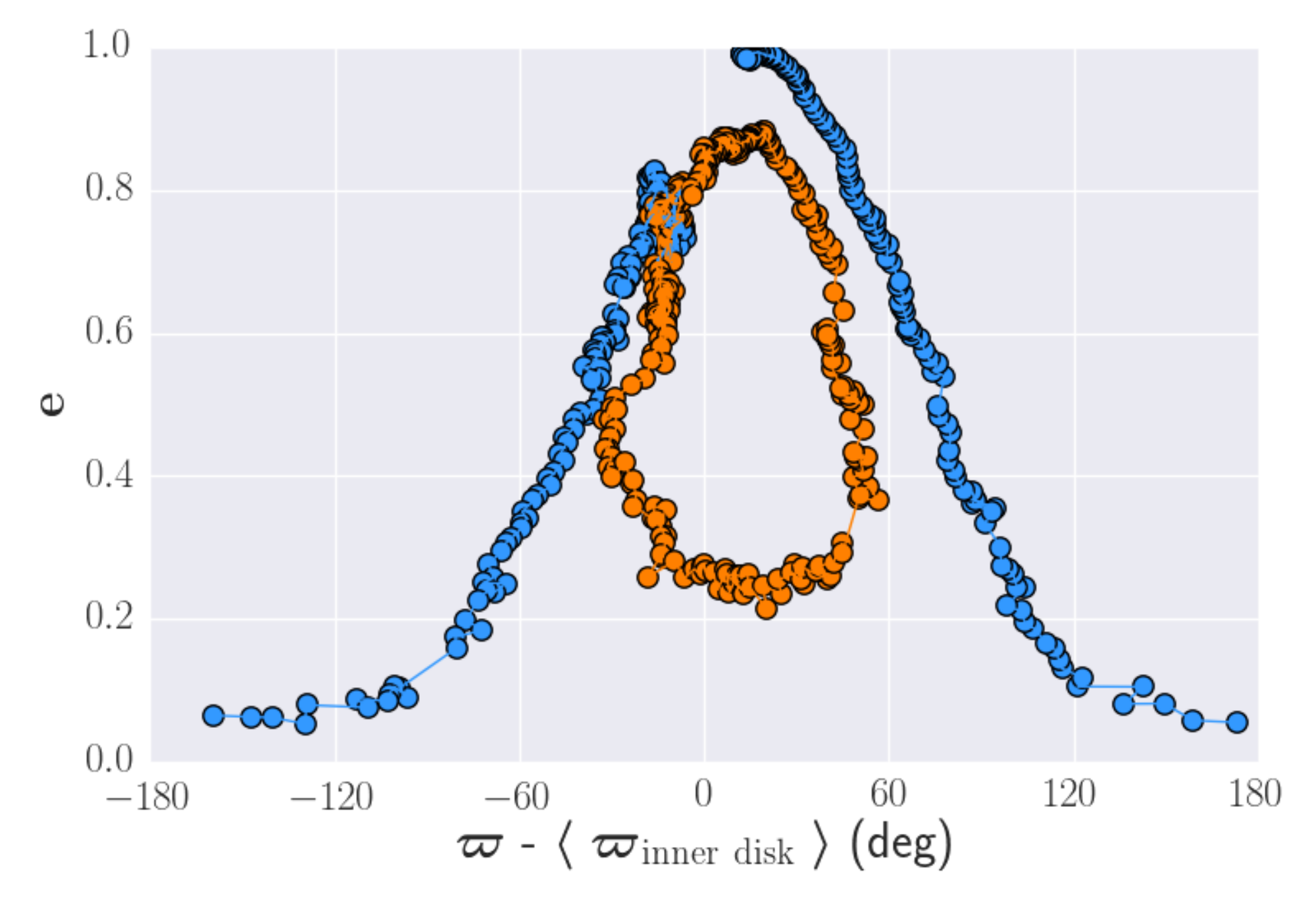}
  \caption{\textbf {Oscillations of orbits}   \\*[0.5\baselineskip]
Eccentricity and longitude of periapsis $e-\varpi$ tracks for two stars in a simulation. Data are plotted every orbit over $\sim300$ orbital periods. The inner (orange) orbit oscillates about the main body of the disk and varies in eccentricity as predicted from our simple model. We subtract the mean longitude of periapsis of the inner disk to highlight this. The outer (blue) orbit demonstrates the exchange of angular momentum between the inner and outer disk as they precess past each other. 
}
 \label{fig:pomega_e}
\end{figure}

\begin{figure*}
  \begin{minipage}{\textwidth}  
    \centering
  \includegraphics[width=1.0\textwidth]{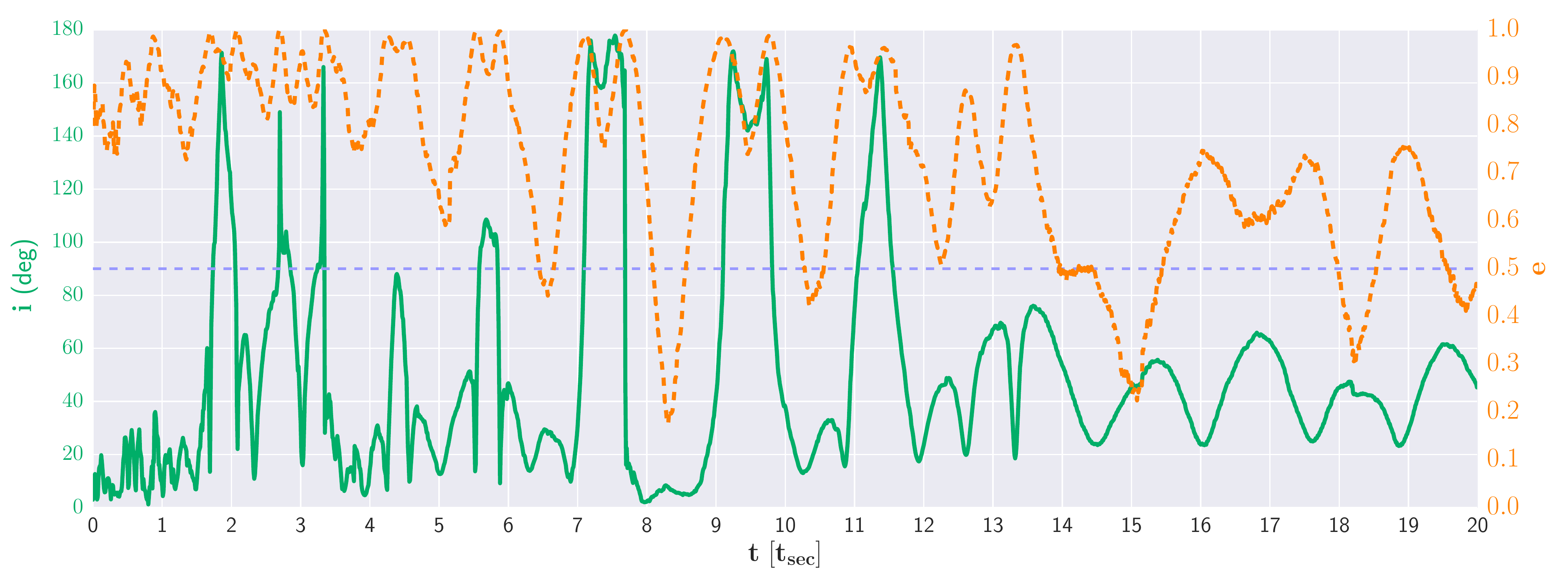}
\end{minipage} \quad

\begin{minipage}{\textwidth}  
    \centering
              \includegraphics[width=0.49\textwidth]{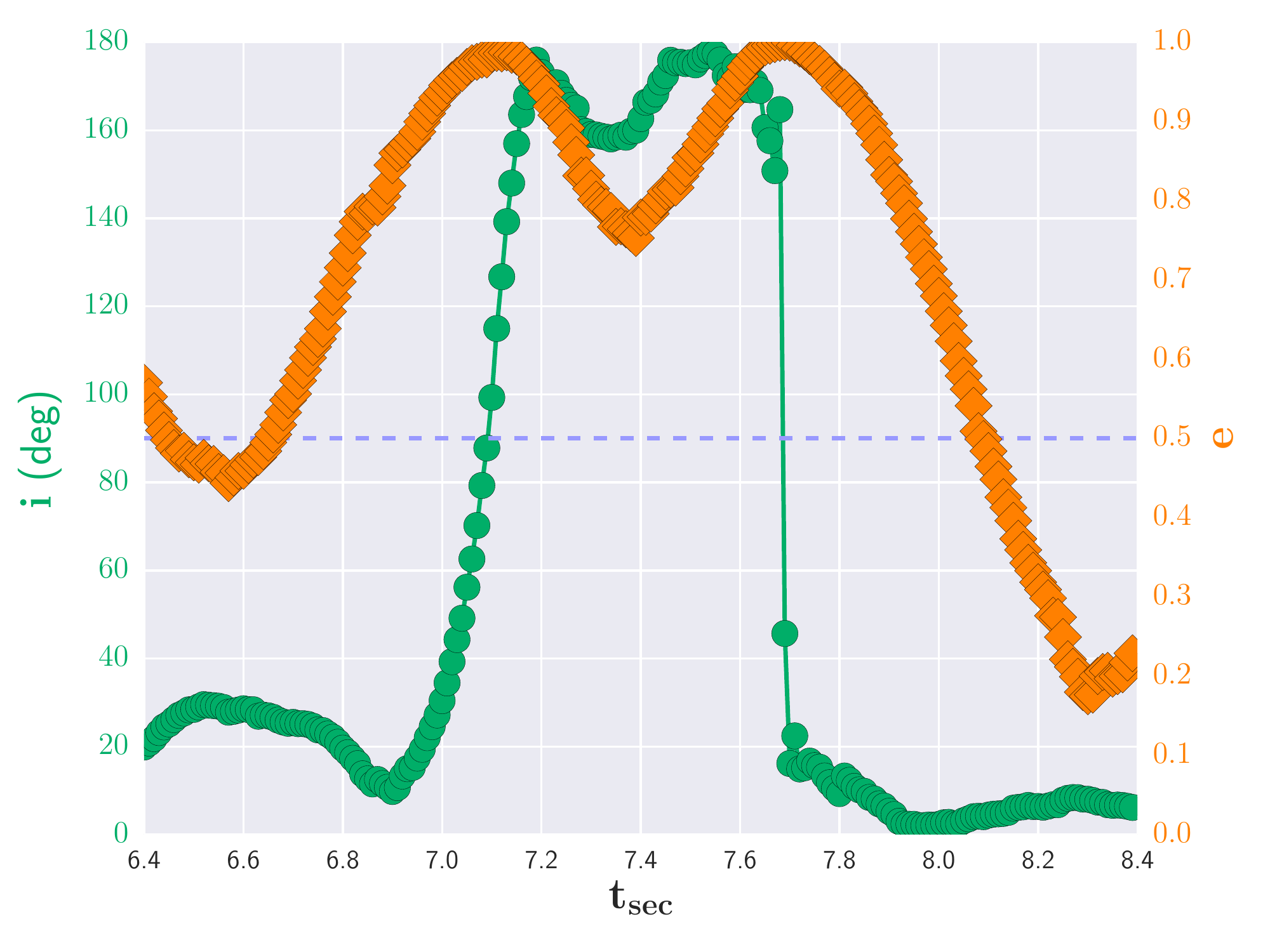}
                    \hspace{0.00\linewidth}
	 \includegraphics[width=0.49\textwidth]{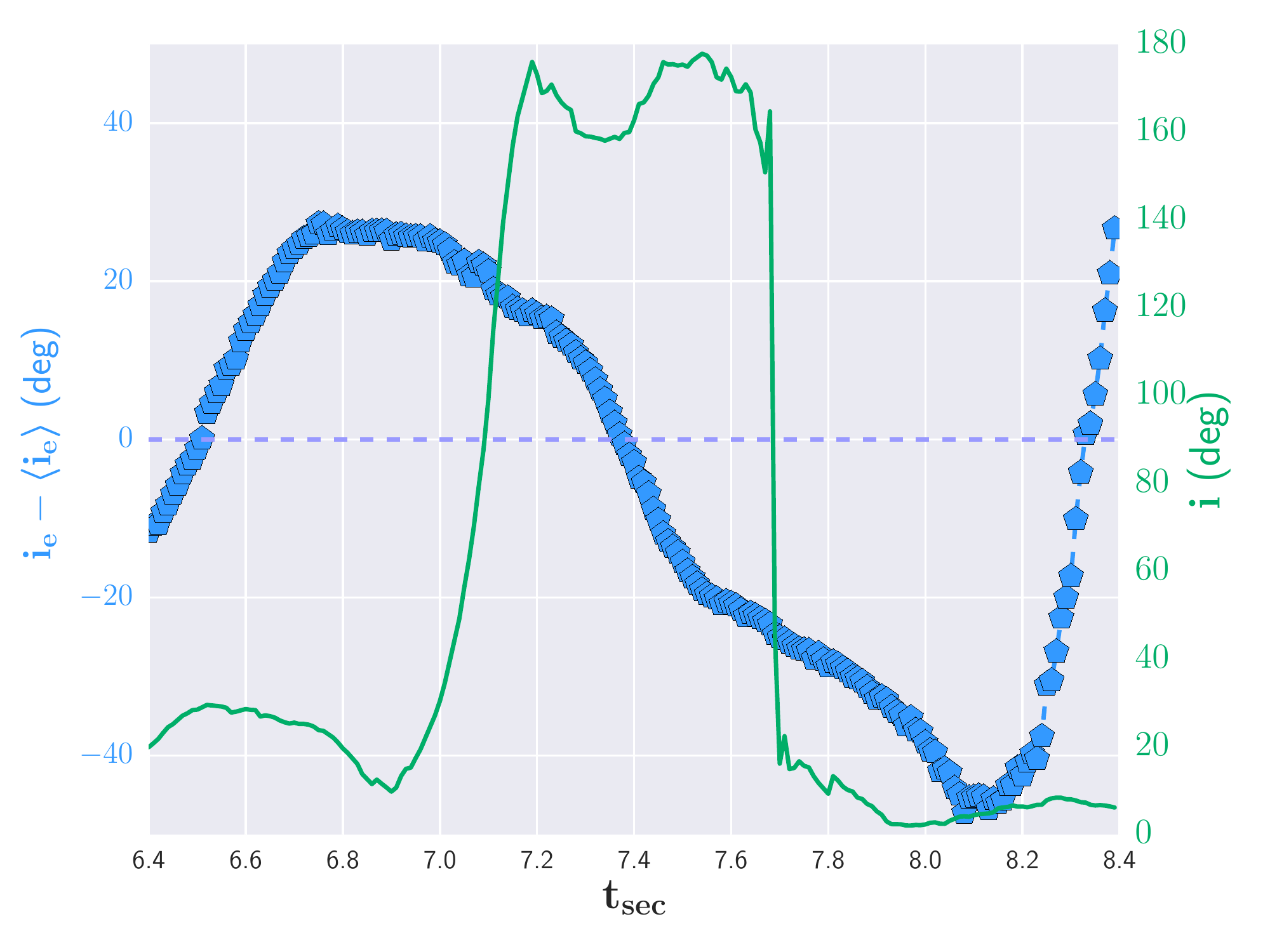}
            \end{minipage}
\caption{\textbf{Inclination and eccentricity of a single star in the stably-precessing inner disk.} \\*[0.5\baselineskip]
(\textit{Top}): Inclination `flips' (green) correspond to high eccentricity peaks (orange). Evolution is shown as a function of time in units of the secular timescale $t_{\rm sec} \equiv  {\Mbh}/{\Md}~ P$. 
(\textit{Bottom}): Zoom-in on inclination flip (green circles, plotted every orbital period), with eccentricity (left; orange diamonds) and relative $\ie (\equiv \arctan{e_y/e_x})$ value with respect to the main body of the inner disk (right; blue pentagons). 
}
\label{fig:flips}
  \end{figure*}

As discussed in \S~\ref{sec:model}, the stabilizing forces which hold the disk together against differential precession also drive oscillations in the individual orbits eccentricity vectors. These oscillations are driven by gravitational torques -- which adjust orbital eccentricities and limit the differential precession rate -- and are damped by two-body relaxation. 
In figure~\ref{fig:pomega_e} we show longitude of periapsis $\varpi$ tracks for two stars over $\sim300$ orbital periods, subtracting the mean longitude of periapsis  of the main disk to highlight orbital oscillations. 
The inner (orange) orbit oscillates with counter-clockwise motion about the main body of the disk as described in \S~\ref{sec:model}. When ahead of the disk ($\varpi - \langle \varpi\rangle > 0$), it increases in eccentricity and decreases in precession rate. When the disk overtakes it ($\varpi - \langle \varpi\rangle < 0$), it decreases in eccentricity and increases in precession rate. 
The (blue) outer orbit differentially precesses (with retrograde motion) on this timescale. It feels strong gravitational torques from the inner disk and so it oscillates in eccentricity as it encounters and moves past the inner disk. 
Here we write down the governing dynamical equations for the orbital oscillations within the inner disk and derive its timescale.

The orientation of an eccentricity vector in the disk can be described by $\ie~ \equiv \arctan{e_y/e_x}$  \citep{Madigan2016}. We note that this angle is equivalent to the longitude of periapsis $\varpi (= \Omega + \omega)$ for orbits of zero inclination, $\ie = \Omega - \arctan\left[\cos \omega, \cos i \sin \omega\right] + \pi/2$. 
The precession rate, that is the time rate of change of \ie, is
\begin{subequations}\label{eq:ie_prime}
\begin{align}
d\ie' &= - \frac{j f_{\text{r}}}{G M_{\bullet} e} \cos \psi + \frac{\tau_{||} v_{\text{r}}}{G M_{\bullet} e} \left[\frac{2}{e} + \cos \psi \right], \label{eq:ie'1} \\
&\sim (e^{-2} -1)^{1/2} ~2\pi~t_{\rm sec}^{-1}  \label{eq:ie'2} 
  \end{align}
\end{subequations}
where $\psi$ is the true anomaly of the star on its orbit, $\tau_{||} \equiv \vec{\tau}\cdot\hat{j}$ is the perpendicular component of the torque and $v_r$ is the radial velocity of the star at $\psi$ \citep{Madigan2017}. 
The first term in equation~\ref{eq:ie'1} describes precession due to a radial, non-Keplerian specific force \fr.
For simplicity we evaluate this at apoapsis\footnote{Note that this approximation is valid only for high eccentricity orbits. Furthermore, the rough approximation of the radial force, $f_r$, which we also made in \S~\ref{ss:ecc-gradient}, does not account for the radial spread in orbits in the disk. 
}
 ($\cos\psi = -1$) and ignore the second term which results from fitting Kepler elements to a slightly non-Keplerian orbit. 
We approximate the magnitude of the \textit{radial} force experienced at apoapsis as $f_r  \sim {G \Md}/{a^2}$ and define the secular dynamical timescale
\begin{equation}
t_{\rm sec} \equiv  \left( \frac{\Mbh}{\Md} \right) P
\end{equation}

We denote the angular difference between the $\vec{e}$-vector of a test orbit and the mean $\vec{e}$-vector of the disk by $\delta \ie$. 
To first order, differences in eccentricity drive spreading of the disk according to
\begin{equation}
\label{eq:linear}
\delta\ie' = \frac{d\ie'}{de} \delta e
\end{equation}
From equation~\ref{eq:ie'2},
\begin{equation}
\frac{d\ie'}{de} = -\left(\frac{e^{-3}} {\sqrt{e^{-2} -1}}\right)~\left(\frac{2\pi}{t_{\rm sec}}\right) \label{eq:die_de}
\end{equation}

Next we model the azimuthal \textit{in-plane} force felt by a test orbit by the disk. The eccentric disk has some mean azimuthal  \note{spread (angular spread of the eccentricity vectors in the disk plane)} which we denote as $\phi_{\rm disk}$. A test orbit lying outside this range experiences a force $f \sim - G\Md / (a \delta \ie)^2$. Inside the disk, however, the force should not tend to infinity. In fact opposing forces $\sim$cancel near the middle of the disk. Here we approximate the force as $f \sim - \alpha \delta\ie$, where $\alpha = G\Md/(a^2 \phi_{\rm disk}^3)$. 
The time rate of change of eccentricity of a test orbit within the disk is given by
\begin{subequations}
\label{eq:e_prime}
\begin{align}
e' &= \frac{2E({\mathbf j}\cdot{\mathbf \tau})}{(G\Mbh)^2e} + j^2 \frac{(\mathbf{v} \cdot \mathbf{f}) }{(G\Mbh)^2e} \label{eq:e'1}\\
&= \frac{\sqrt{e^{-2} -1}}{\phi_{\rm disk}^3} ~\delta\ie ~\left(\frac{2\pi}{t_{\rm sec}}\right)
  \end{align}
\end{subequations}
where we have ignored the second half of the equation~\ref{eq:e'1} as it oscillates over an orbital period, and used $\tau \sim r f \cos(\delta \ie) \sim rf$. 

Differentiating equation~\ref{eq:linear}, and equating $\delta e' = e'$, gives 
\begin{subequations} \label{eq:die''_de}
\begin{align}
{\delta\ie''} &= \frac{d\ie'}{de} e' \\
&= - (e \phi_{\rm disk})^{-3} ~\delta\ie ~\left(\frac{2\pi}{t_{\rm sec}}\right)^{2} 
  \end{align}
\end{subequations}
This yields the equation for simple harmonic motion, $\delta\ie'' = - \omega_{\rm osc}^2 \delta\ie$, with the oscillation frequency 
\begin{equation}
\label{eq:omega}
\omega_{\rm osc} = (e \phi_{\rm disk})^{-3/2} ~\left(\frac{2\pi}{t_{\rm sec}}\right). 
\end{equation}
The oscillation period is 
\begin{equation}
\label{eq:t_osc}
t _{\rm osc}= \frac{2\pi}{\omega_{\rm osc}} = (e \phi_{\rm disk})^{3/2} ~t_{\rm sec}
\end{equation}
Taking representative values of $e \sim \phi_{\rm disk} \sim 0.8$, the oscillation period in our simulations should be $O$(100 P). 
This is the order of magnitude we see for oscillations in figures~\ref{fig:pomega_e}, ~\ref{fig:flips}, and ~\ref{fig:e_v_t}.
  
\subsection{Inclination flips} 
\label{sec:flips}

Since stars with very high orbital eccentricity have low angular momentum, they are easily torqued around and may undergo orbital `flips'. 
In figure~\ref{fig:flips} we focus on one such star. The top plot shows its eccentricity (orange; right $y$-axis) and inclination (green; left $y$-axis) evolution over the entire simulation. 
High eccentricity peaks often correspond to a rapid reversal of orbital inclination. 
The bottom two plots show a zoom-in at a time when the orbital inclination (green circles, plotted every orbital period) flips from prograde to retrograde and back in tandem with eccentricity oscillations (orange diamonds). In the bottom right plot,  we show $\ie (\equiv \arctan{e_y/e_x})$ relative to the mean $\ie$ of the inner disk (blue pentagons), essentially quantifying the difference in eccentricity vectors in the plane of the disk. 
The double-peaked eccentricity profile and corresponding double inclination flip can be explained using simple Newtonian mechanics as follows: 

As the orbit precesses ahead of the main body of the disk ($t \sim 6.5$), its eccentricity increases due to the disk's negative gravitational torque. Its precession rate slows to a halt with respect to that of the disk ($t \sim 6.8$), and the coherent torque efficiently reduces the orbit's angular momentum until it approaches and passes through zero. The inclination is flipped to $i \sim 180\deg$ and the torque is now acting in the same direction as the angular momentum vector of the orbit and so starts to circularize it. As the disk overtakes the orbit ($t \sim 7.4$), the torque reverses direction and yet again the eccentricity increases. In the same manner as before, the torque drives the angular momentum to zero and flips its direction. The inclination returns to prograde in a matter of orbits. 

These rapid inclination flips are not new. 
\citet{Tou09} observe similar behavior in simulations of nearly-coplanar, counter-rotating orbits, using a softened Gauss code which secularly evolves systems of gravitationally interacting Kepler ellipses. Eccentricities oscillate from near-circular to near-radial and back, as inclinations flip between prograde and retrograde near the peaks of the eccentricity oscillation. 
These flips have also been detected in $N$-body simulations of mutually-interacting eccentric orbits and disks \citep{Loc09b,Kazandjian2013,Haas2016,Subr2016}.

\citet{Li2014a} study similar near-coplanar flips in the context of hierarchical three-body systems using Hamiltonian dynamics. They identify the dynamics with the octupole eccentric Kozai-Lidov  mechanism (third order in semi-major axis ratio between inner and outer binary). 
If we relate the outer perturber in their models with the coherent eccentric disk in ours, it is unsurprising that we find similar dynamics. \note{We note however, that the high eccentricities and similar semi-major axes of eccentric nuclear disk stars fail the stability criterion for hierarchical three-body systems }
\begin{equation}
\epsilon = \frac{a_1}{a_2} \frac{e_2}{1 - e_2^2} < 0.1
\end{equation}
\note{where the subscripts $1, 2$ refer to the inner and outer orbit respectively \citep[][for a review]{Naoz2016}. In the eccentric nuclear disk scenario, $a_1 \approx a_2$, $e_2 \gtrsim 0.4$, $\epsilon \gtrsim 0.5$. }

\note{
Since these orbital flips occur only when the stars reach very high eccentricity, and consequently reach small periapsis distance, one might expect the high number of inclination flips to decrease when we include general relativity in our $N$-body simulations.
General relativistic precession acts in a prograde direction at a rate which increases with orbital eccentricity. This increases the differential precession rate between perturbed, high-eccentricity orbits and the disk. Preliminary results suggest however that the rate of inclination flips remain constant; secular torques from the eccentric nuclear disk are strong enough to push orbits through zero angular momentum faster than general relativistic precession can respond (Wernke et al, in prep.). }

\section{Enhanced rate of TDEs}
\label{sec:tdes}

A Tidal Disruption Event (TDE) occurs when a star is destroyed by the tidal force of a supermassive black hole \citep{Hil75b}. A significant fraction of the disrupted stellar gas remains bound to the supermassive black hole \citep{Ree88}, and its subsequent accretion produces a luminous flare. Dozens of TDE candidates have been observed to date in X-ray, UV and optical wavelengths \citep[for review see][]{Komossa2015}. Here we propose that eccentric nuclear disks may be one of the dominant sources of TDEs in the universe. 

We have shown that a stable eccentric nuclear disk develops a negative eccentricity gradient. Consequently, a star's distance of closest approach to the black hole, $p = {a(1-e)}$, strongly decreases with decreasing distance to the black hole. The eccentricity oscillations which occur on top of this negative eccentricity gradient means that low semi-major axis stars at the peak of their eccentricity oscillation are vulnerable to the tidal effects of the SMBH at periapsis. 

\begin{figure}
  \centering
  \includegraphics[width=1.0\columnwidth]{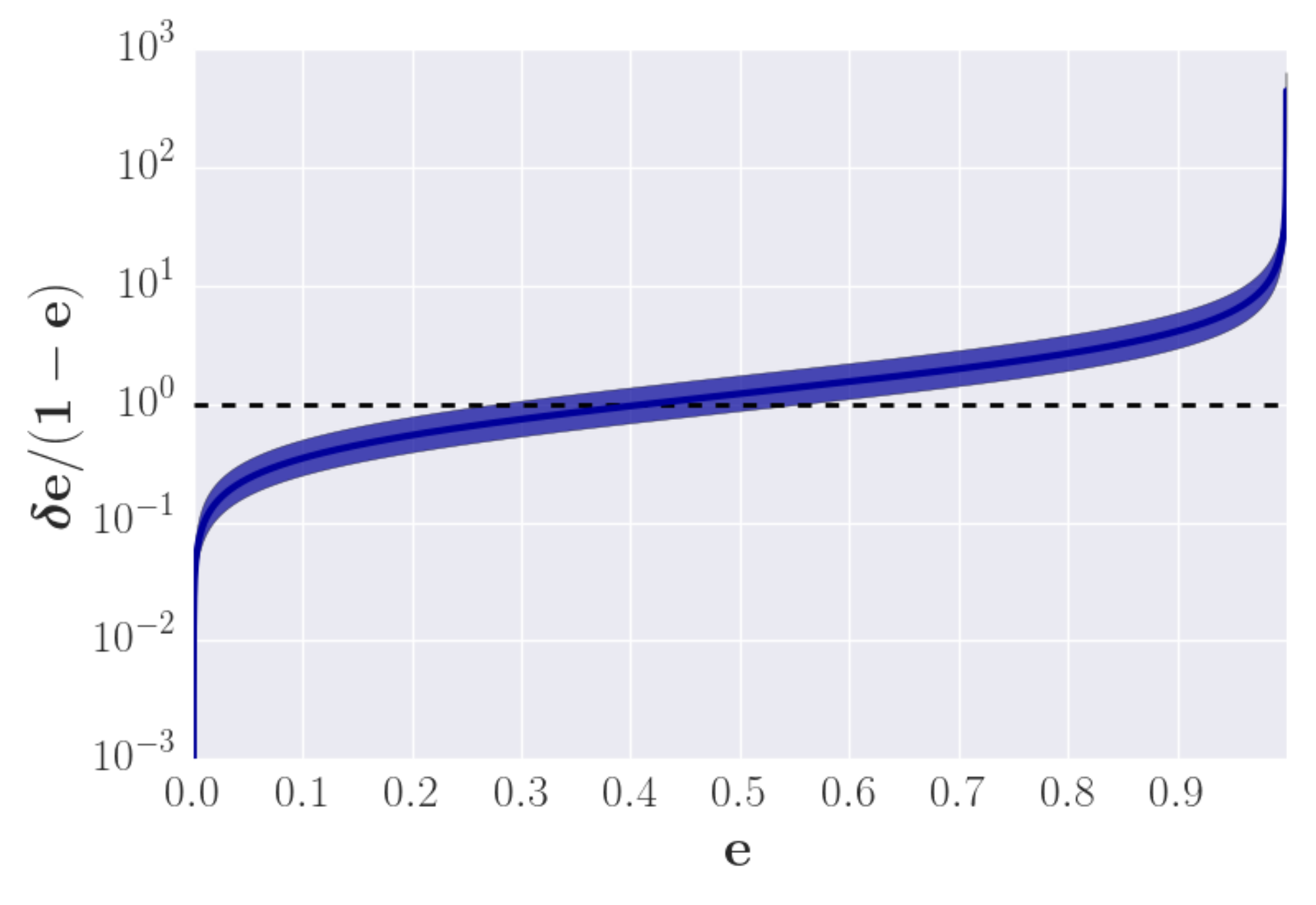}
  \caption{\textbf {TDE Threshold}   \\*[0.5\baselineskip]
 A star is in danger of being disrupted by the SMBH when its equilibrium eccentricity plus its perturbed eccentricity tends to one, ${\delta e}/(1 - e) \to 1$ (equation~\ref{eq:switch}). We plot this as a function of equilibrium eccentricity. 
  At high eccentricities, it is a very steep function (note the logarithmic axis). 
 Stars with equilibrium eccentricities $e \gtrsim 0.6$ may result in a TDE.  
 As the mean eccentricity of an eccentric nuclear disk decreases with time (as stellar mass, but not angular momentum, is lost to the SMBH), the stellar orbits move to the left of this plot until they drop below the threshold, ${\delta e}/(1 - e) = 1$, and the disk no longer produces TDEs. 
 }
 \label{fig:tde_switch}
\end{figure}

\begin{figure*}
  \centering
  \includegraphics[width=1.0\textwidth]{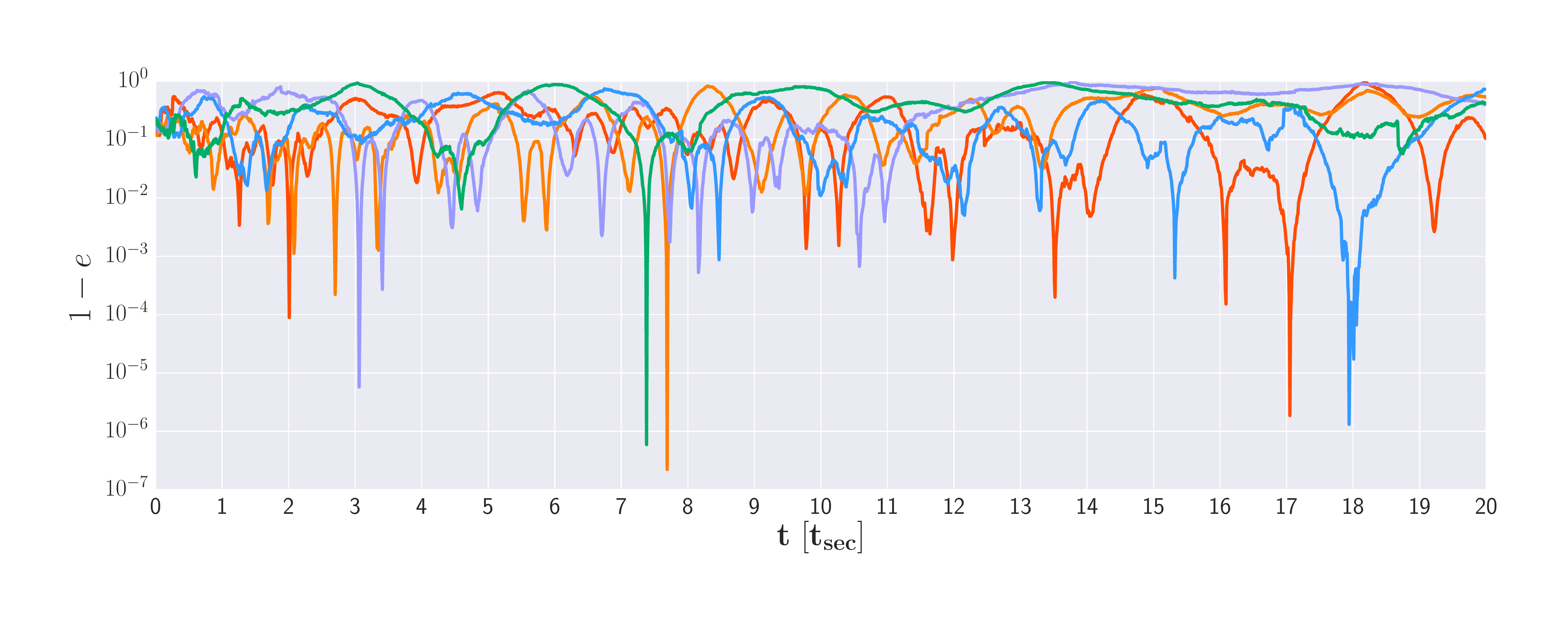}
  \caption{\textbf {Eccentricity evolution of stars in the inner prograde-precessing disk.}   \\*[0.5\baselineskip]
Orbital eccentricities of stars in an $N$-body simulation with initial disk eccentricity of $e = 0.8$ as a function of time in units of the secular timescale $t_{\rm sec} \equiv  {\Mbh}/{\Md}~ P$ ($= 100 P$). The widths of the eccentricity peaks are often narrow, just a few orbits wide. Though we select these stars for their high eccentricities, we note that $36\%$ of stars in this simulation reach $1 - e < 10^{-4}$. 
}
 \label{fig:e_v_t}
\end{figure*}

Stars will disrupt if, during their orbital oscillations, their eccentricities approach values of unity ($ e + \delta e \to 1$). 
From our linear analysis of the stabilizing forces in the previous section, we can derive a ``TDE threshold'' for oscillations: 
\begin{equation}
\label{eq:switch}
\frac{\delta e}{1 - e}  \sim  \sqrt{\frac{e~(1+ e)}{(1 -e)}} ~ \phi_{\rm disk}^{-3/2} ~ \delta\ie
\end{equation}
where, as before, $e$ is the initial eccentricity of the orbit, $\delta e$ is the magnitude of the eccentricity oscillation, and $\phi_{\rm disk}$ quantifies the angular ``width'' of the disk in radians, comparable to the amplitude of oscillation ($ \delta\ie$). 
The disk rapidly produces TDEs whenever the TDE threshold ${\delta e}/(1-e) \gtrsim 1$.  We plot equation~\ref{eq:switch} as a function of eccentricity $e$ in figure~\ref{fig:tde_switch}, taking a range of values for $\phi_{\rm disk}$ and $\delta\ie$: $0.5 - 2$ radians. 
The TDE threshold is met whenever the oscillating orbits at the inner edge of the disk have equilibrium (un-perturbed) eccentricities $e{\gtrsim}0.6$, and those with  $e \gtrsim 0.9$ greatly surpass it. 
Figure~\ref{fig:e_v_t} demonstrates this, showing the eccentricity evolution of a handful of stars with low semi-major axes (and hence large equilibrium eccentricities) in an $N$-body simulation with initial disk eccentricity $e  = 0.8$. Time is given in units of the secular timescale. Over the course of $\sim$two precession periods of the inner disk, all five stars are torqued to $(1 - e) < 10^{-5}$. 

Since the mechanism stabilizing the disk as a whole sets up a large-scale eccentricity gradient, simply disrupting the innermost stars will not deplete the supply: they will be dynamically repopulated by differential precession on approximately an oscillation timescale.  This is the secular-dynamical analogue of ``refilling the loss cone'', and it is \textit{much} faster than the diffusive processes typically considered when computing predicted TDE rates \citep[e.g.,][]{Wan04,Stone2016a}. 
However, if we consider an eccentric nuclear disk in isolation, we see that such an elevated rate of TDEs cannot persist indefinitely. Since each TDE removes mass but not angular momentum from the disk, it lowers the mean eccentricity of the disk.  Eventually the disk will drop below the TDE threshold ${\delta e}/({1-e}) = 1$, causing the TDE rate to dwindle. This will also cease evolution of the disk, so we predict most eccentric disks to have inner edges with $e \sim 0.6 - 0.7$.  
On the other hand, if the eccentric nuclear disk is embedded in a star cluster and its potential is an attractor, background stars may join the structure. This could replenish the disk's mass as the innermost stars are lost to the SMBH, thus prolonging the phase of elevated TDE rates. 

In our $N$-body simulations and in figure~\ref{fig:tde_switch}, we observe that the inner edge of the disk needs to drop below $e \sim 0.6$ before the TDEs switch off. 
{This should not be taken as a precise prediction. Both the calculations leading to equation~\ref{eq:switch} and our $N$-body simulations are overly-simplistic. Nevertheless, the fact that they agree with one another is encouraging.}
Given the negative eccentricity gradient, an inner edge of $e \sim 0.6$ corresponds to a mean disk eccentricity of $\langle e \rangle \sim 0.4$. If the mean disk eccentricity begins at $\langle e \rangle \sim 0.8$, this requires a third of the disk mass to drain onto the SMBH. Until then, the self-gravity of the disk continually forces stars into the SMBH.

 We cannot currently make accurate estimates of the time-dependent TDE rate from eccentric nuclear disks using our $N$-body calculations. 
 As the two-body relaxation rate scales as $\sim N^{-1}$, dissipation in our low-$N$ simulations is artificially high. Convergence studies with high-$N$ simulations are computationally demanding and beyond the scope of this work.  
However, we expect the effect of dissipation is to lower the amplitude of the oscillations and thus to artificially reduce the TDE rates in our simulations. 
 Additionally, we will need to improve our calculations with the addition of bulge gravitational potentials and apsidal precession due to general relativity (although general relativistic precession does not appear to lower TDE rates; Wernke et al, in prep.). 

Though we cannot yet use $N$-body simulations to calculate the TDE rate, we can make an order of magnitude estimate for the TDE soon after disk formation. We take a representative SMBH mass of $\Mbh = 10^7 ~\Mo$ and an SMBH-disk mass ratio similar to what is observed in M31 ($\sim10\%$), to give a disk mass of $\Md = 10^6~\Mo$. 
We estimate that $\sim10\%$ of stars are disrupted soon after disk formation\footnote{In simulations with mean initial disk eccentricity of $e = 0.8$ ($e = 0.6$), over a third ($15\%$) of the stars reach $(1 - e) < 10^{-4}$ at some point during the first 2000 orbital periods.} . We make the simplest approximation and convert this directly to $10^5$ stars. The TDEs won't happen instantaneously; we estimate that they occur over ten oscillation times, which yields $10^5$ stars being disrupted over $10^2 - 10^3$ orbital periods (equation~\ref{eq:t_osc}). 
The stars are preferentially disrupted at the inner edge of the disk. From observations of the M31 nuclear eccentric disk and the (non-apsidally aligned) nuclear disk in the Galactic center \citep{Lu2009}, we estimate the inner disk radius to be $\sim10^{-2}~r_H~ \approx$ 0.05 pc, where $r_H$ is the radius of influence of the SMBH. The orbital period at this radius is $\sim300$ yr. Hence, we arrive at a rate of $0.3 - 3~\peryr~\pergal$.
We emphasize that this is an extremely crude estimate. Nevertheless it illustrates the ease at which high TDE rates can be generated by an eccentric nuclear disk. 
There is essentially no classical angular momentum barrier, since self-gravity of the disk actively forces stars into the black hole. 

Our model anticipates anomalously high TDE rates from gaseous, recently-merged galaxies \citep{Hop10a,Hop10b}.   
TDEs at this stage may be difficult to distinguish from quasar or AGN activity.  
Certainly, overlapping events will obscure the characteristic TDE luminosity signature \citep[$L \sim t^{-5/3}$;][]{Phinney89}. 
TDEs interacting with a pre-existing magnetized accretion disk, built up from previous TDEs, may account for observed jetted TDEs \citep{Tchekhovskoy2014,Kelley2014}. 
Radiation pressure in hyperaccreting TDE envelopes may also produce jets in TDEs \citep{Coughlin2014}, and significant variability can be induced in the fallback rates through gravitational fragmentation of the tidal debris stream \citep{Coughlin2015}. 
TDEs involving high-metallicity stars \citep{Batra2014,Kochanek2016} may help explain observed abundance anomalies in quasars. 

\citet{Tadhunter2017} recently discovered the first TDE candidate in a nearby ultra-luminous infrared galaxy (ULIRG). ULIRGs represent the peak of major, gas-rich galaxy mergers. Unlike in most ULIRGs, the central region of this galaxy is not heavily enshrouded in dust and so presents us with a rare view of its star-forming, active nucleus.  Given this unusually clear sightline and the small sample of ULIRGs from which the event was discovered, the authors suggest that the TDE rate in such galaxies could be $\sim0.1~\pergal~\peryr$. Interestingly, the TDE flare was unusually prolonged compared with typical TDEs, with the light curve flattening at late times rather than following the typical $L \sim t^{-5/3}$ decline predicted for isolated events.

Since the eccentric disk loses stellar mass but not angular momentum through TDEs, its mean eccentricity will decrease with time. Lowering the eccentricity at the inner edge of the disk reduces the probability of TDEs, as shown in figure~\ref{fig:tde_switch}. Hence the TDE rate will decrease with time. Eventually, individual events will be detectable on top of the baseline AGN activity. This may in some cases explain the so-called ``changing-look quasars'' \citep{Macleod2016}, in which transient broad H$\alpha$ emission lines, possibly due to TDE flares, are seen superimposed on quasar spectra. 

When the time between TDEs is longer than the accretion time of the stellar material onto the SMBH ($\sim 10^{-2}~\pergal~\peryr$), TDEs will be individually identifiable.  Thus we predict that TDEs occur at a lower rate in post-merger galaxies (with the rate continuing to decrease with time since merger) than in ULIRGs, but that events should behave more like typical TDEs following the theoretical $L \sim t^{-5/3}$ decline. 
There may already be evidence to support this: 
\citet{Arcavi2014} and \citet{French2016} have discovered that TDEs frequently occur in K+A galaxies.  
K+A (or E+A) galaxies are so called due to Balmer absorption features in their spectra (characteristic of  an A star) which appear superimposed on an old K star or (E)arly-type galaxy population \citep{Dressler1983}.  
The Balmer absorption in K+A galaxies points to a significant starburst population with ages $\sim10^8-10^9$ years, while low H$\alpha$ indicates a lack of ongoing star formation. There is strong evidence that galaxy-galaxy interactions and/or mergers trigger the starburst \citep{Yang2004}, and the most luminous K+A galaxies appear to be successors of ULIRGs.
There is even a connection to the formation of eccentric nuclear disks: 
the population of A-stars in K+A galaxies is often very centrally concentrated, a consequence of gas driven to the center during the merger \citep[e.g.,][]{Yang2008}, which hydrodynamic simulations of galaxy formation suggest leads to eccentric disks \citep{Hop10a, Hop10b}. 
The results of \citet{French2016} imply an average TDE rate of $\sim 10^{-3}~\peryr$
per K+A galaxy, and only a few $\times~10^{-6}~\peryr$ per
galaxy in `normal' star-forming or elliptical galaxies.  

In our model, when the inner edge of the disk drops to $e \lesssim 0.6$, the eccentricity oscillations are inefficient at driving stars to $e\sim1$. Tidal interactions are greatly reduced and so the eccentric disk structure `fossilizes'. The double nucleus in M31 is an example of such a remnant disk. 

\section{Discussion}
\label{sec:disc}

This paper focuses on the dynamics of \textit{eccentric nuclear
 disks}, in which the stars orbiting the super-massive black hole in
the center of a galaxy all have eccentric, apsidally-aligned orbits.
In an eccentric nuclear disk, the individual stellar orbits nearly
overlap, such that the entire star cluster takes the form of an
elliptical disk with the black hole at one focus.  While such a
configuration at first seems exceedingly improbable, an eccentric
nuclear disk has been observed in the nearby galaxy M31, and quite
possibly in many other galaxies as well.  Eccentric nuclear disks are
only discernible in very nearby galaxies, and double nuclei appear only for nearly edge-on
configurations; the low probability of detecting eccentric nuclear
disks therefore suggests they may be quite common in the local
universe.  It is surprising that such an apparently finely-tuned
structure should be seen in so many galaxies, and this result suggests
that some unknown process actively creates and stabilizes eccentric
disks.

While self-gravity seems like a perfect candidate for stabilizing
eccentric disks, ostensibly one would expect self-gravity to
instead \textit{de}stabilize these disks since it drives differential
precession.  Stellar orbits precess in the non-keplerian potential
caused by the disk's gravity; since stars near the inside and outside
of the disk feel different gravitational potentials, they should
precess at different rates, smearing out the angular extent of the
disk.

We have shown that such smearing by differential precession does not
in fact occur; the same non-keplerian forces which drive precession
also torque the orbits and change their eccentricities.  Since the
precession rate also depends on eccentricity ($\propto \vec{f} \times \vec{j}/e$), we find
the system reaches a stable equilibrium in which the orbital energies
and angular momenta of individual stars balance such that each star
has a nearly identical precession rate, independent of its location
within the disk.  We sketch the dynamics of this
mechanism in figure~\ref{fig:model}, and we show that the system is stable in the
sense that stars undergo oscillations about their equilibrium
configuration (see figure~\ref{fig:pomega_e}).  Our model explains the $\gtrsim$\,Gyr
longevity of the M31 disk, and its negative eccentricity gradient.

In addition, our model makes a number of predictions, which may be
testable in M31 and other nearby galaxies:
\begin{enumerate}
\item We predict the presence of \note{a less massive ($\le 10\%$) outer disk} in the M31 nucleus (\S~\ref{ss:outer-disk}),
 which precesses in the opposite direction to the stable eccentric nuclear disk.  Preliminary results from ``observing" our $N$-body  simulations, indicates that the inner and outer disk can reproduce M31's double nucleus structure. 
 
\item In addition to the negative eccentricity gradient in M31, we
 also predict a negative inclination gradient (\S~\ref{ss:ecc-gradient}). 
 
\item Our model predicts a small population of counter-orbiting stars at low semi-major axes due to rapid inclination flips (\S~\ref{sec:flips}). 

\item Our stability model predicts that eccentric disks may be quite common in
 nearby galaxies, especially post-merger galaxies.  These disks
 will only manifest as double-nuclei over a narrow range of
 parameter space, but may appear as offset nuclei or as a central
 ``hole'' in the stellar distribution \citep{Lauer2005}.

\end{enumerate}

\subsection{Implications for Galaxy Formation}
Using cosmological simulations, \citet{Hop10a, Hop10b}, show that
tidal torques can drive gas into galactic nuclei during galaxy
mergers.  If enough gas enters the galactic nucleus, it forms a
lopsided, eccentric disk orbiting the supermassive black hole.  When
this gas fragments and forms stars, it may produce an eccentric
stellar disk very similar to the ones we study here.

Not all eccentric disks are stable; reversing the logic in figure~\ref{fig:model},
one can see that if the orbits precess in a retrograde manner, the
disk becomes unstable.  Retrograde precession would naturally result from a massive 
symmetric background stellar population, such as the one found in the
Milky Way Galactic Center.  This may explain why we do not observe one
in our own Galaxy, despite the observational evidence for a recent
star formation event in a disk \citep{Lev03}.  However, we note that
during a galaxy merger, in-spiraling SMBHs are expected to scour out a
core in stars in any prior nuclear star distribution leading to the
conditions necessary for stability of a stable eccentric disk.

Once a stable eccentric disk forms, it should persist until either the
background potential changes, or until the galaxy undergoes another
merger.  In either event, eccentric nuclear disks may last for many
Gyr, serving as persistent markers of gas-rich galaxy mergers.  The
population of nearby eccentric disks may therefore encode valuable
information about galaxy mergers and evolution in the local universe.

\subsection{Implications for Tidal Disruption Events}
Eccentric nuclear disks are stable in the sense that they undergo
stable oscillations about their equilibrium configuration.  Over the
course of these oscillations, individual stars swing in both
eccentricity and in orientation.  Since the equilibrium configuration
of the disk places the innermost stars at the highest eccentricity,
these oscillations may take the innermost stars to an eccentricity of
one if the overall disk eccentricity is high enough.  Such stars are on plunging orbits and are therefore
susceptible to tidal disruption by the supermassive black hole.  This
leads to the corollary that eccentric nuclear disks may be remarkably
prolific sources of TDEs: the self-gravity of the
disk actively funnels stars into the black hole, with essentially no
analogue of the loss cone.  This process should continue until the
stellar disk partially circularizes, likely after losing a significant
fraction of its mass to TDEs.

While we cannot yet precisely quantify either the total number of
TDEs from a single eccentric nuclear disk, or precisely how the TDE rate evolves with time, we can describe the
expected behavior in a broad qualitative manner:
\begin{itemize}
\item At early times the TDE rate should be very high ($0.3 -
 3~\peryr~\pergal$) and nuclei will be difficult to distinguish
 from an AGN.  Along many lines of sight, TDEs will be blocked by
 the eccentric disk itself.
\item As stars are destroyed through tidal forces, the disk loses
 mass though not angular momentum and so should become less
 eccentric over time.  Since a lower initial eccentricity decreases
 the number of stars susceptible to disruption, the TDE rate will
 decrease with time, unless the mass of the disk is somehow
 replenished.
\item When the rate has dropped sufficiently such that the average
 TDE duration is shorter than time period between TDE events ($\sim
 10^2~\peryr~\pergal$), individual TDEs will be easier to identify.
 At this point a galaxy may be in the post-merger phase (E+A/K+A).
\item When the inner edge of the disk drops to $e{\lesssim}0.6$, the
 eccentricity oscillations are inefficient at driving stars to
 $e\sim1$.  Tidal interactions are greatly reduced, the disk ceases
 to evolve, and the eccentric disk `fossilizes.'  This state may
 persist for many Gyr.
\end{itemize}

\subsection{TDE Predictions}
If our hypothesis is correct, and eccentric nuclear disks are a dominant source of TDEs, gas-rich mergers and K+A/E+A galaxies should host eccentric stellar disks within the radius of influence of their SMBHs. Unfortunately both of these galaxy types are rare and typically too distant to resolve nuclear scales. 
When unresolved, an eccentric nuclear disk will appear as an asymmetric nucleus, as in the case of M31 before the era of {\textit HST}. 
\citet{Stone2016b} recently demonstrated using HST photometry that one of our nearest E+A galaxies, NGC 3156 ($22$ Mpc), has a significantly asymmetric nucleus. 

Our model predicts a high rate of TDEs originating from a recently-formed eccentric nuclear disk. Therefore we expect the age of stars being tidally disrupted to increase as a function of time since merger. Recent major, gas-rich mergers or (U)LIRGs should host TDEs of the youngest stars, with K+A/E+A galaxies hosting an older ($\sim0.1-1$ Gyr) stellar TDE. We expect massive galaxies to host more TDEs than smaller galaxies, up until the SMBH mass at which the tidal disruption radius lies within the Schwarzschild radius ($\Mbh \lesssim 10^8 \Mo$). 
Eccentric nuclear disks orbiting more massive SMBHs will still produce high eccentricity orbits and in-falling stars will add to the growth of the SMBH, but they will not result in observable TDEs.

The mechanism driving stars to high eccentricities works via secular gravitational torques, not two-body relaxation. Therefore we might expect some stars to repeatedly graze the loss cone before fully disrupting. 
Partial TDEs and TDEs with stripped outer envelopes should be commonly observed. The latter should lack hydrogen in their spectra \citep[\eg PS1-10jh;][]{Strubbe2015}. 
Multiple stellar systems, if they prevail in eccentric nuclear disks, will be more easily disrupted than individual stars. This could result in hyper-velocity stars, binaries and tightly-bound high eccentricity stars close to the SMBH, detracting somewhat from the overall TDE rate.  
\note{Other likely outcomes include double tidal disruptions and disruptions of highly magnetized stars  
\citep{Mandel2015,Bradnick2017}.}

\subsection{Future work}

We will study the structure of eccentric nuclear disks as a function of initial disk parameters such as eccentricity and surface density. We will also explore how the disk responds to background stellar potentials and losing stellar mass via TDEs. Ultimately we would like to calculate the time-dependent TDE rate due to evolving eccentric nuclear disks. We will include the effects of general relativity, background gravitational potentials, and stellar mass segregation.
 
With a suite of $N$-body simulations of differing initial conditions, we will be able to calculate the photometric and spectroscopic signatures of eccentric nuclear disks at different resolutions. These may be used to determine the prevalence of eccentric nuclear disks in the local universe, and directly compared with high-resolution studies of the nuclear environments of ULIRGS \citep[e.g.,][]{Medling2014} and E+A/K+A galaxies. 
Since eccentric nuclear disks are long-lived, it's also worth considering their observational implications; in particular, how their presence may bias SMBH mass measurements.

We will additionally study the effect that the eccentric nuclear disk has on the background stellar population surrounding the SMBH. Torques from the non-axisymmetric disk potential, analogous to the eccentric Kozai-Lidov mechanism \citep{Li2015,Naoz2016}, should produce high eccentricity orbits in this older stellar population. This will result in an older TDE population along with hyper-velocity stars and hyper-velocity binaries. We are interested to explore whether the latter can account for the currently unexplained distribution of unusual Ca-rich supernovae observed offset from their host galaxies \citep{Perets2010,Kasliwal2012,Lyman2014,Foley2015}. 
Finally, torques due to eccentric nuclear disks are an exciting possible channel for the production of extreme-mass ratio inspirals (EMRIs) of compact stellar remnants into supermassive black holes, a prime target for low-frequency, space-based, gravitational wave observatories such as \textit{LISA}. 

\section*{Acknowledgments}

\noindent{} AM thanks Mariska Kriek, Marcel van Daalen, and Freeke van de Voort for their encouragement at UC Berkeley. She thanks Decker French, Ann Zabludoff and Nadia Zakamska for enlightening conversations. 
 AM was supported by UC Berkeley's Theoretical Astrophysics Center. 
 MkMcC was supported by NASA grant NNX15AK81G and HST-HF2-51376.001-A, under NASA
contract NAS5-26555. 
 CN was supported by the Science and Technology Facilities Council  (STFC) (grant number ST/M005917/1). 
 \note{We gratefully acknowledge support from NASA Astrophysics Theory Program (ATP) under grant 16-ATP16-0114.}
 Resources were provided by the NASA High-End Computing (HEC) Program through the NASA Advanced Supercomputing (NAS) Division at Ames Research Center under grant SMD-15-6582.

\bibliographystyle{apj}

\end{document}